\documentclass[traditabstract]{aa}
\usepackage{longtable}
\usepackage{graphicx,natbib,txfonts,color}
\bibpunct{(}{)}{;}{a}{}{,}

\begin{document}

\title{HERMES: a high-resolution fibre-fed spectrograph for the Mercator telescope
\thanks{Based on observations made with the Mercator Telescope,
  operated on the island of La Palma by the Flemish Community, at the
  Spanish Observatorio del Roque de los Muchachos of the Instituto de
  Astrof\'{i}sica de Canarias. }
}

\author{Gert Raskin\inst{1}
\and Hans Van Winckel\inst{1}
\and Herman Hensberge\inst{2}
\and Alain Jorissen\inst{3}
\and Holger Lehmann\inst{4}
\and Christoffel Waelkens\inst{1}
\and Gerardo Avila \inst{5}
\and Jean-Pierre De Cuyper \inst{2}
\and Pieter Degroote \inst{1}
\and Ren\'e Dubosson \inst{6}
\and Louis Dumortier \inst{2}
\and Yves Fr\'emat \inst{2}
\and Uwe Laux \inst{4}
\and Bernard Michaud \inst{6}
\and Johan Morren \inst{7}
\and Jesus Perez Padilla \inst{1}
\and Wim Pessemier\inst{1}
\and Saskia Prins\inst{1}
\and Kristof Smolders \inst{1}
\and Sophie Van Eck\inst{3}
\and Johannes Winkler \inst{4}
}

\offprints{G. Raskin, Gert.Raskin@ster.kuleuven.be}

\institute{ Instituut voor Sterrenkunde, K.U.Leuven, Celestijnenlaan 200D,
B-3001 Leuven, Belgium 
\and  Royal Observatory of Belgium, Ringlaan 3, 1180 Brussel, Belgium
\and  Institut d'Astronomie et d'Astrophysique, Universit\'e Libre de
Bruxelles, CP226, Boulevard du Triomphe, B-1050 Brussels, Belgium 
\and Th\"uringer Landessternwarte Tautenburg, Sternwarte 5, D-07778
Tautenburg, Germany
\and European Southern Observatory, Karl-Schwarzschild-Str. 2, 85748 Garching, Germany
\and Observatoire de Gen\`eve, Chemin des Maillettes 51, CH-1290 Sauverny, Switzerland
\and Dept. of Physics and Astronomy, K.U.Leuven, Celestijnenlaan 200D,
B-3001 Leuven, Belgium
}

\date{Received 20 July 2010 / Accepted 11 October 2010}

\authorrunning{G. Raskin et al.}
\titlerunning{HERMES}

\abstract{
The HERMES high-resolution spectrograph project aims at exploiting the specific
potential of small but flexible telescopes in observational
astrophysics. The optimised optical
design of the spectrograph is based on the well-proven concept of 
white-pupil beam folding for high-resolution spectroscopy. 
In this contribution we present the complete project, including the
spectrograph design and procurement details, the telescope adaptor and 
calibration unit, the detector system, as well as the 
optimised data-reduction pipeline. 
We present a detailed performance analysis
to show that the spectrograph
performs as specified both in optical quality and in total
efficiency. With a spectral resolution of 85\,000 (63\,000 for the low-resolution fibre),
a spectral coverage from 377 to 900\,nm in a single exposure and a peak
efficiency of 28\%, HERMES proves to be an ideal instrument for
building up time series of high-quality data of variable (stellar) phenomena.}

\maketitle

\section{Introduction}\label{sect:intro}

The Mercator telescope (\texttt{www.mercator.iac.es}) is a \mbox{1.2-m}
optical telescope based at the Roque de los Muchachos Observatory on La Palma
(Canary Islands, Spain) and funded by the Flemish community of
Belgium. The telescope is operated by the Instituut voor Sterrenkunde
of the K.U.Leuven university (Belgium).
The telescope itself is a twin telescope of the Swiss 
Euler Telescope operational at the La Silla observatory of the
European Southern Observatory (ESO).  Both telescopes are ideal
complements to international facilities. Their specific niche in
observational astrophysics is that they guarantee long
and secured access, which allows for dedicated monitoring programmes
of variable phenomena.

To exploit the telescope infrastructure, a focused
instrument development programme was deployed and in this contribution
we present HERMES, an acronym for 
high efficiency and resolution Mercator echelle spectrograph.
This instrument was realised by a consortium of institutes with
main science drivers in stellar astrophysics. The project was led by
the Instituut voor Sterrenkunde (K.U.Leuven, Belgium) with as main 
partners the Institut d'Astronomie et Astrophysique (ULB,
Belgium) and the Royal Observatory of Belgium, and with smaller
contributions from the Th\"uringer Landessternwarte Tautenburg (Germany) and the Geneva
Observatory (Switzerland).

The science drivers cover a wide range of topics in stellar
astrophysics that mainly call for dedicated high-quality time series
of high-resolution spectra. They comprise the
ground-based follow-up of asteroseismology space missions and
spectroscopic asteroseismology studies in general
\citep[e.g.][]{aerts10}, massive (interacting) binaries
\citep[e.g.][]{hensberge08, lehmann10, blomme09}, final phases of
stellar evolution in (binary) stars
\citep[e.g.][]{vanwinckel09, jorissen09}, and
photospheric abundance determinations as probes to internal
nucleosynthesis
\citep[e.g.][]{masseron10, reyniers07a}.
This resulted in a series of science requirements that cover
spectrograph characteristics (spectral resolution $>$ 80\,000; spectral range $>$ 380\,--\,880\,nm;
throughput $>$ 25\% in V, cycle time $<$ 60\,s; radial velocity stability $<$ 5\,m/s), 
as well as operational goals (flexible but robust night operations). 
The last are important, given the limited local staff and night support at the Mercator telescope.
To accommodate these requirements, a fibre-fed spectrograph with two science modes was designed: one optimised for high spectral resolution with good wavelength stability and excellent throughput, and the other optimised for very high  stability with lower resolution and throughput.

In this contribution we present the HERMES spectrograph project, which
was realised in between the kick-off meeting of 19 January 2005 and the
start of the regular science operations in April 2009. We start with
a description of the instrument design in Section \ref{sect:design} and of the 
data-reduction pipeline in Section \ref{sect:drs}. In Section \ref{sect:performance} we 
report on the performance analyses, followed by conclusions in Section \ref{sect:conclusion}.

\section{Instrument design}\label{sect:design}

\subsection{Spectrograph}

\subsubsection{Optical design}
\begin{figure*}
\centering
\includegraphics[width=16cm]{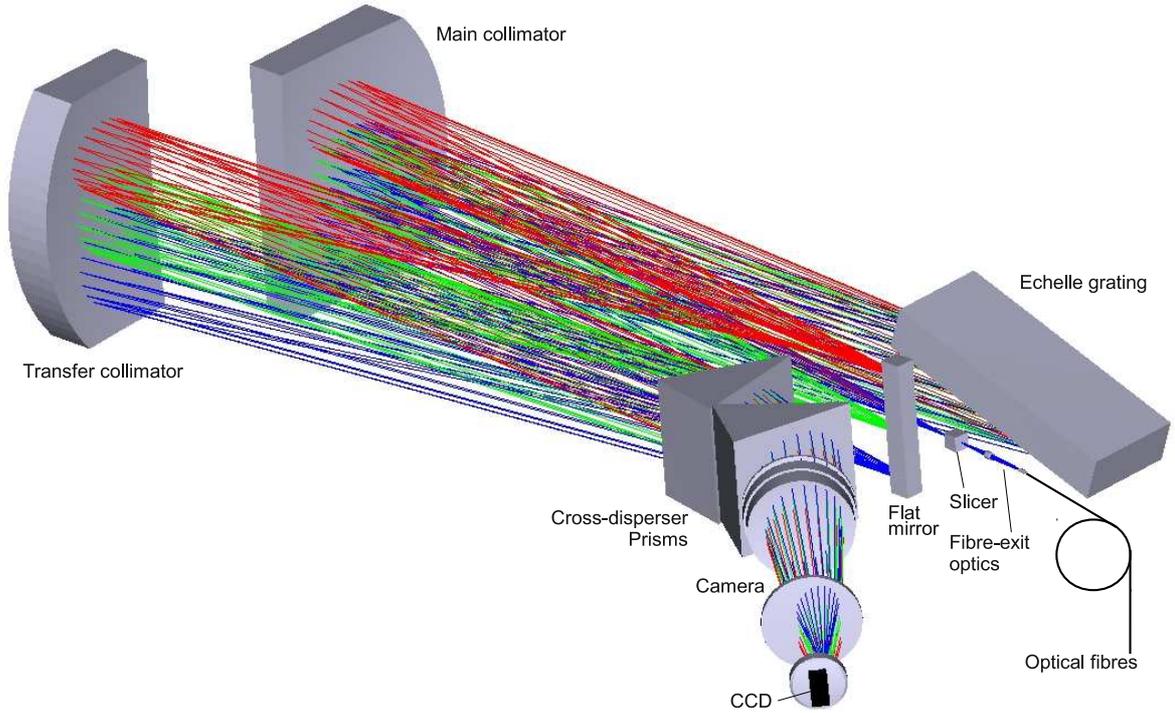}
\caption{\label{fig:layout}3D view of the HERMES optical layout.}
\end{figure*}

Like many present-day high-resolution spectrographs, the optical
layout of HERMES (Fig.~\ref{fig:layout}) is based on a white-pupil
design, discussed in detail by \citet{Baranne88}. This concept
implemented with a pair of off-axis parabolas as collimator, as 
first done for UVES \citep{Dekker00}, makes it arguably one of the best
solutions for a high-resolution spectrograph.  It was also used for
other successful ESO spectrographs like FEROS \citep{Kaufer98},
generally accepted as a benchmark for instrument efficiency, and HARPS
\citep{Pepe00} as a benchmark for radial-velocity precision. These projects became
important inspirations for HERMES, which attempted to
combine the strengths of these instruments. These instruments are
also fibre-fed, and the HERMES design follows their well-proven
concept.

\paragraph{Echelle grating}
HERMES uses an echelle grating blazed with an intermediate blaze angle
of $\theta$\,=\,$69.74^{\circ}$ ($\tan\theta$\,=\,2.7) as main dispersing
element. The ruling of 52.676 grooves/mm nicely distributes the
spectrum from 377\,nm to 900\,nm in 55 spectral orders over the
rectangular CCD. The aluminium-coated ruled surface on a Zerodur blank
measures 154\,x\,408\,mm. The grating is used in Littrow condition, but is
slightly tilted ($\gamma=0.8^{\circ}$) perpendicular to the
direction of dispersion to separate the dispersed from the incident
beam. To avoid dust contamination, the grating is mounted with the
optical surface facing downwards. The efficiency of the grating at
blaze peak amounts to 65\%. 

\paragraph{Collimators} 
The HERMES collimators were cut from a single
660-mm diameter Zerodur parabolic mirror with a focal length of 1400\,mm
(Seso, France). With an aperture of $f/9.2$, the collimated beam
has a diameter of 152\,mm. The mirrors are coated with a blue-enhanced
protected silver coating (Spectrum\,Thin\,Films, USA) with very good
reflectivity in the blue and green parts of the spectrum ($\mbox{Refl.}>96\%$
at 400\,nm, $\mbox{Refl.}>99\%$ over 450 -- 650\,nm) but with a somewhat
disappointing and below-specification performance in the
near-infrared ($\mbox{Refl.}>94\%$ at 800\,nm).

\paragraph{Cross disperser}
Two identical prisms used at minimum deviation for the spectrograph's
central wavelength (490\,nm) separate the spatially overlapping
echelle orders. Thanks to the white-pupil mounting, the size of the
cross-disperser prisms could be minimised, but they still are two
impressive blocks of glass with a base of 138\,x\,208\,mm, a height of
197\,mm, and an apex angle of $37.4^{\circ}$. Both prisms are made of
Ohara I-Line PBL1Y glass with very high blue and UV transmission
(Optique Fichou, France). 
The cross disperser provides a minimum order separation of 24 pixels in
the reddest part of the spectrum, a value that increases to 55 pixels
at 380\,nm. The inter-order distance is sufficiently large to record
two interleaved spectra in case of the simultaneous thorium technique
\citep{Baranne96}, separated by 13 pixels on the CCD.

The ratio between the longest and shortest wavelength is
2.4 for HERMES. The rejection of unwanted orders over such a broad spectral
range is complicated when using a grating-type cross disperser. Moreover, the prism solution has higher efficiency (20-30\%), avoids the production of cross-order scattered light, and offers more
evenly spaced orders. We therefore discarded the use of a grating as in UVES \citep{Dekker00} or a grism
as in HARPS \citep{Pepe00}.

\begin{figure}
\resizebox{\hsize}{!}{\includegraphics{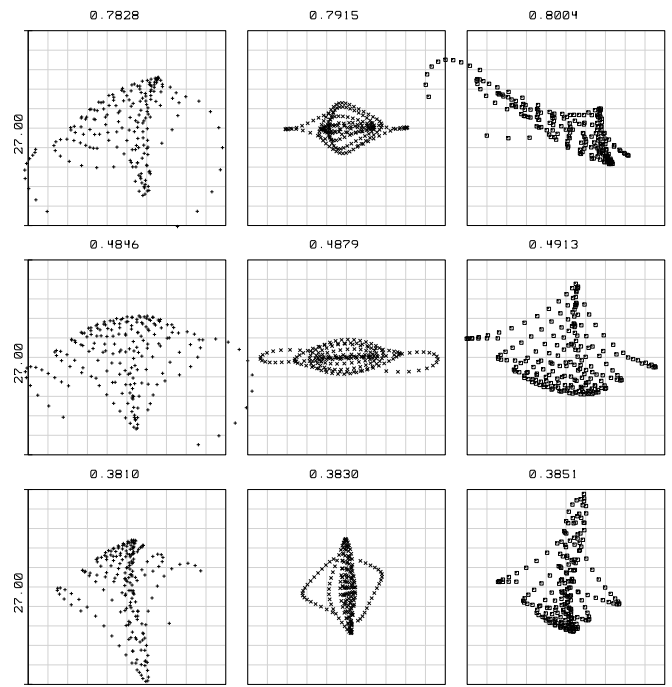}}
\caption{\label{fig:spotdiagrams} HERMES spot diagrams at beginning, centre, and end of the free spectral range for three representative orders (43, 73, and 93). Wavelength is indicated in $\mu$m above each 2 x 2 pixels box (27\,$\mu$m).}
\end{figure}

\paragraph{Dioptric camera}
The camera (Seso, France) that focuses the spectrum on the detector is
a fully dioptric system consisting of three singlet lenses and one
large cemented triplet. The last lens acts as field flattener with a
convex cylindrical rear surface. This lens also serves as a vacuum
window of the detector cryostat. The camera has a focal length of
475\,mm and a focal ratio of $f/3.1$. This gives the spectrograph a
linear dispersion of 0.002\,nm/pixel at the centre of the bluest
order, increasing to 0.0047\,nm/pixel for the reddest order. The
camera characteristics are calculated to image all but the reddest
spectral orders without gaps on the detector. From 838\,nm onwards, 3
orders are slightly truncated. Sacrificing lateral colour correction
completely, the moderate complexity of this camera nonetheless delivers
excellent image quality as can be appreciated from the
spot diagrams for the complete optical system of the spectrograph in
Fig.~\ref{fig:spotdiagrams}. The overall throughput of the camera is
higher than 88\% for most of the wavelength range.

\paragraph{Fibre-exit optics and exposure metre}
The light beam exiting the optical fibres enters the spectrograph at a $f/3.7$ focal ratio. This is converted to a $f/9.2$ beam -- accepted by the collimators -- by a set of focal-ratio-adaptation optics, consisting of one doublet and one triplet (Precision Optics Gera, Germany). The doublet is cemented directly onto the fibres for increased mechanical stability and to limit the number of glass-air interfaces, thereby reducing reflection losses.

\begin{figure}
\resizebox{\hsize}{!}{\includegraphics{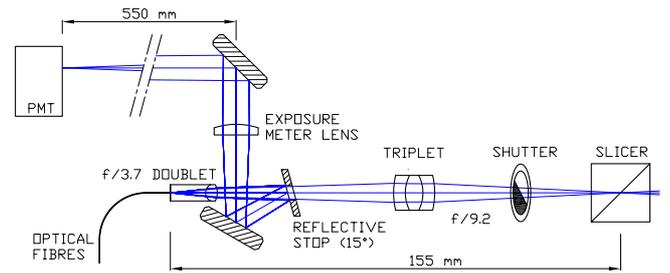}}
\caption{\label{fig:exp_meter} Optical fibres entering the spectrograph, focal-ratio-adaptation optics, exposure metre and slicer.}
\end{figure}

\begin{figure}
\resizebox{\hsize}{!}{\includegraphics{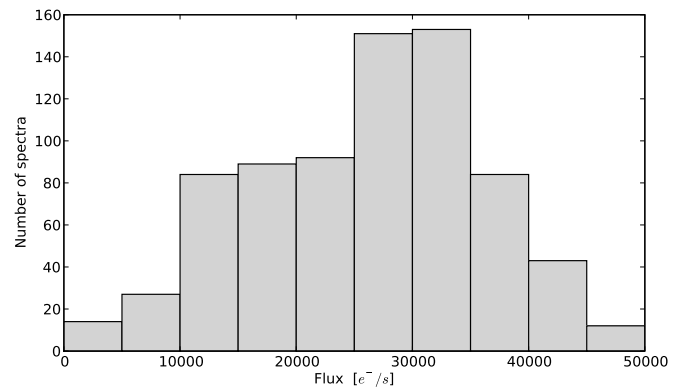}}
\caption{\label{fig:flux_histogram} Histogram of the flux measured in V, normalised to the $0^{th}$ magnitude, for a large sample of HERMES spectra.}
\end{figure}

In between the two lenses and conjugated with the echelle grating, the inclined reflective stop of the adaptation optics collects those photons otherwise lost from the main beam due to focal ratio degradation (see Section \ref{sect:fibre_link}) and sends them to a photo-multiplier tube for continuous flux monitoring. One lens and two fold mirrors image the fibres on the photo-multiplier detector (Fig.~\ref{fig:exp_meter}). The flux from the thorium fibre is masked out to avoid contamination of the stellar flux. This exposure metre allows the determination of accurate photon-weighted mean time of the exposures, necessary to correct for the Earth's velocity. It is also very useful for matching the exposure time with the desired signal-to-noise ratio (S/N) of the final spectrum during observations under poor weather conditions. The broad distribution of the flux histogram in Fig.~\ref{fig:flux_histogram} clearly illustrates the effect of variable observing conditions and the resulting need for continuous flux monitoring.

\begin{figure}
\resizebox{\hsize}{!}{\includegraphics{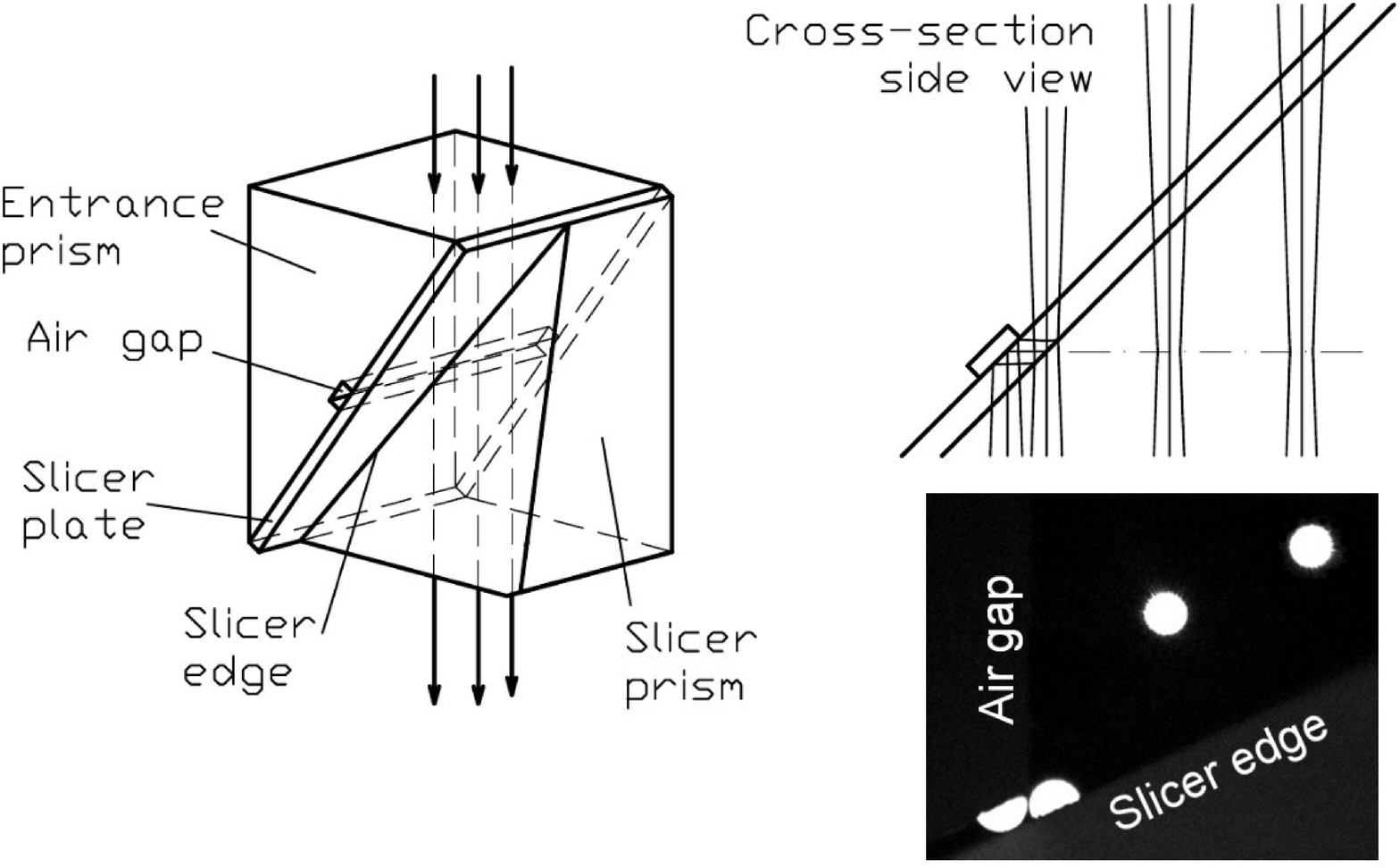}}
\caption{\label{fig:slicer} Perspective view and cross-section of the slicer. The picture shows an image of the simultaneously illuminated sliced and unsliced fibres. Inside the spectrograph, the slicer edge is aligned horizontally.}
\end{figure}

\paragraph{Slicer}
A two-slice image slicer effectively doubles the spectral resolution
in the high-resolution mode. As the cross-order profile of the sliced
fibre is too wide or the inter-order spacing too small to sample two
interleaved and sliced spectra, the slicer cannot be used for
simultaneous thorium exposures. For the same reason a simultaneous
sky fibre for sky substraction was not implemented. 
The image slicer is similar to the FEROS type
\citep{Kaufer98b} but here only one fibre image is sliced. The images
of the two fibres used in the simultaneous thorium mode pass unsliced
through the slicer. The slicer layout and operation are shown in
Fig.~\ref{fig:slicer}.

The design of the slicer allows imaging of the sliced and unsliced
fibres without any focus difference. Only a negligible focus difference 
of 0.25\,mm exists between the two slices of the sliced image. The
slicer consists of an entrance prism with one air groove, the slicer
plate whose thickness (0.173 mm) determines the distance between the
two slices (0.277\,mm), and the cut-off slicer prism. The prism angle
is $46.5^{\circ}$ guaranteeing total internal reflection up to the
longest wavelengths. The sides of the prism measure 25\,mm. All parts
are made from Homosil quartz and connected by optical contact (Horst
Kaufmann, Germany). The entrance and exit surfaces of the slicer are
coated with a broadband AR coating. The very high throughput of the
slicer (above 95\%) largely compensates for the increased complexity
compared to the use of a slit for obtaining higher resolution.

\subsubsection{Mechanical design}
Mechanical stability is a main issue in the design of a spectrograph for measuring precise radial velocities. HERMES is therefore mounted on a vibration-absorbing optical bench (2400\,x\,1200\,x\,300\,mm). For more mechanical stability, this bench is located on a solid concrete foundation, isolated from the rest of the telescope building. As the spectrograph works in a completely fixed configuration, there are no moving parts on the optical bench that might compromise this stability with the exception of a tiny shutter, installed at the fibre exit where the beam diameter is minimal (Fig.~\ref{fig:exp_meter}). All large optical elements have (semi-)kinematic mountings that constrain all six degrees of freedom independently and without redundancy.

\begin{figure}
\resizebox{\hsize}{!}{\includegraphics{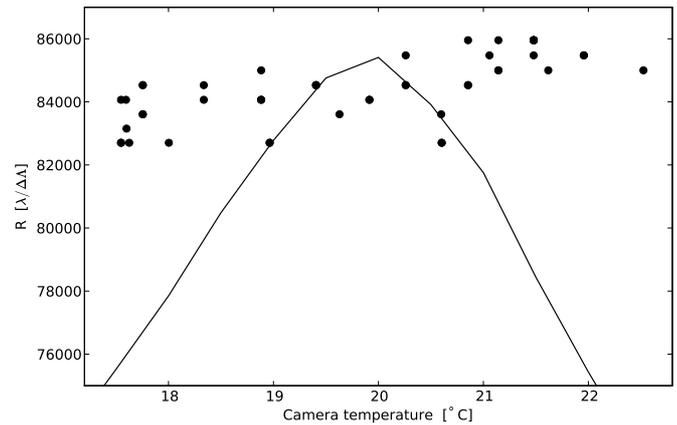}}
\caption{\label{fig:focus_thermal_sensitivity} Effect of temperature on spectral resolution or spectrograph focus (dots). The theoretical thermal sensitivity based on a ZEMAX model without compensation is shown as a solid line for comparison.}
\end{figure}

\begin{figure}
\resizebox{\hsize}{!}{\includegraphics[angle=90]{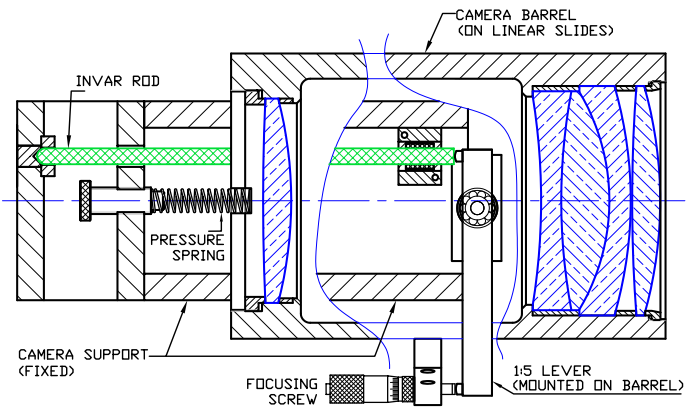}}
\caption{\label{fig:camera} Camera focusing mechanism with temperature compensation.}
\end{figure}

\begin{figure}
\resizebox{\hsize}{!}{\includegraphics{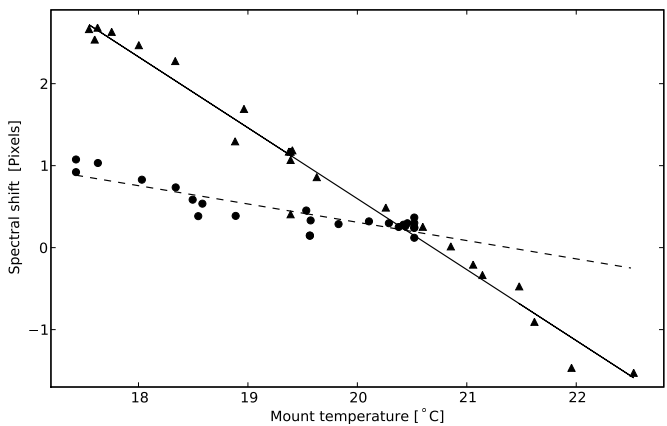}}
\caption{\label{fig:thermal_sensitivity} Measured thermal shift in spectral direction with (dots, dashed line) and without (triangles, solid line) thermal compensators installed in the grating mount.}
\end{figure}

Although the temperature of the spectrograph room is precisely controlled (see Section \ref{sect:env_control}), special care was taken with the thermal stability of the opto-mechanics as well. Most of the mounts for the optical elements are manufactured from cast iron because of its excellent resistance to deformation and, more importantly, its coefficient of thermal expansion (CTE), which matches the CTE of the ferromagnetic stainless steel of the spectrograph bench quite well. Moreover, the mounts of the most critical optics incorporate some type of passive thermal stabilisation to reduce the sensitivity to temperature variations:
\begin{itemize}
    \item Thermal expansion of both the metal structures of the spectrograph and the lenses (mostly in the camera), as well as thermal changes in the refractive index of air and lens glasses, have a non-negligible effect on the spectrograph focus. Temperature changes greater than $1^{\circ}$C would require a focus correction or would result in resolution loss. The focusing mechanism of the camera is therefore equipped with a thermal compensation system that maintains the spectrograph in focus over a broad temperature range. Fig.~\ref{fig:focus_thermal_sensitivity} illustrates the insensitivity of the spectrograph focus to temperature variations  of $\pm3^{\circ}$C, together with the expected defocus without compensation calculated by ZEMAX. Thermal compensation is based on a 520-mm long Invar rod that pushes against the camera barrel through a lever. The opposite side of the lever holds the focusing screw (Fig.~\ref{fig:camera}). Differential thermal expansion of the Invar rod and the cast iron camera support, amplified by the 1:5 lever, keep the spectrograph well-focused independently of temperature.
    \item The cryostat support and the detector support inside the cryostat are designed to have their centre of thermal expansion exactly on the optical axis of the spectrograph camera.
    \item The vertical position of the fibre-exit unit is precisely kept at the focal point of the Zerodur parabolas through a tuned combination of Invar and brass in the fibre support.
    \item Similarly, the angular position of the grating is thermally
      neutralised by the differential thermal expansion of Invar and
      stainless steel in the kinematic grating
      supports. Figure~\ref{fig:thermal_sensitivity} shows the
      fourfold reduction in the thermal sensitivity when using
      thermally compensated kinematic supports. Imperfect modeling
      left us with a drift of 0.22 pixels/$^{\circ}$C, but this can
      be further reduced empirically.
\end{itemize}

\subsubsection{Environment control}\label{sect:env_control}
Precise control of the spectrograph environment is critical for the long-term stability of the instrument, with temperature and air pressure the most important parameters. Fluctuations in the relative humidity (RH) are less relevant as they have less influence on the refractive index of air. In any case, the air conditioning system keeps the spectrograph air very dry with RH$_{avg}$\,=\,13\% and RH$_{max}$\,$<$\,30\%. The higher outliers correspond with physical interventions in the spectrograph room during commissioning and will disappear in the future.

\paragraph{Temperature}
A triple isolation of the spectrograph is the basis of precise thermal control. The spectrograph bench is completely packed by a box of demountable 60-mm panels of polyurethane sandwiched between thin steel sheets. Inside this box, thermal dissipation is minimal and limited to the photo multiplier of the exposure metre ($\sim$250 mW) and the actuator for the miniature shutter. The latter is a bistable solenoid that only dissipates for a few milliseconds during opening or closing of the shutter. The inner room, built from the same 60-mm isolating panels, surrounds the spectrograph box. A LakeShore Model331 temperature controller with 50-Watt heater actively stabilises the temperature of this room at $18\pm0.01^{\circ}$C. Finally, similar 100-mm walls for the outer room isolate the spectrograph further from the outside world. An air conditioning system keeps the temperature in this room at $14\pm0.1^{\circ}$C.

\begin{figure}
\resizebox{\hsize}{!}{\includegraphics{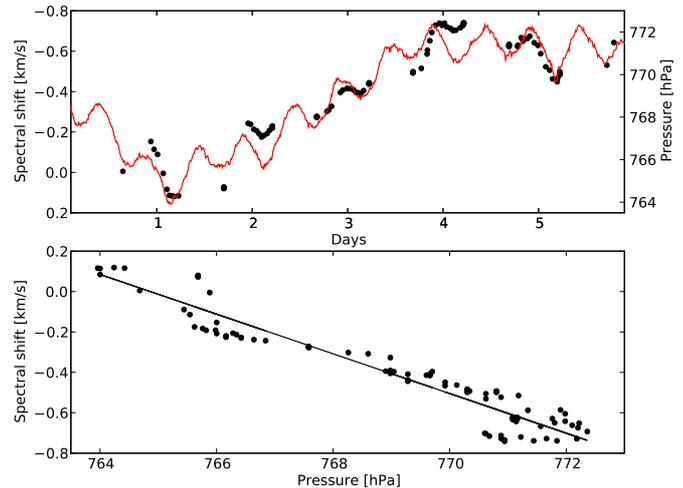}}
\caption{\label{fig:pressure} Radial velocity shift (dots) and typical atmospheric pressure fluctuations (red line) over several days.}
\end{figure}

\paragraph{Pressure}
Apart of temperature, atmospheric pressure changes also have an effect on the refractive index of air, causing a spectral shift. Fig.~\ref{fig:pressure} shows typical pressure fluctuations including the \mbox{(semi-)}diurnal atmospheric tide \citep{Lindzen79}, together with the resulting radial velocity offset that amounts to 0.05 pixel/hPa or 0.08 km/(s\,hPa).

Housing the spectrograph in a vacuum tank would have been the obvious approach for avoiding pressure-induced instrumental drift. However, practical and financial limitations prohibited this solution. As an alternative, we have started the installation of a system that should stabilise the spectrograph at a constant absolute pressure of $\sim$\,782\,hPa, a value close to the maximum atmospheric pressure that occurs at the altitude of the observatory. To allow for over-pressure of maximum $\sim$\,20\,hPa in the spectrograph room during periods of low atmospheric pressure (observing conditions never occur when atmospheric pressure is below 762\,hPa), the walls of the outer room have been reinforced by a structure of 40-mm steel tubes.

\subsection{Optical fibre link}\label{sect:fibre_link}
The  large dimensions of HERMES prohibit the installation of the
instrument at the telescope focal plane. The obvious solution is
to use optical fibres to relay the light from the telescope to the
spectrograph. Aside from causing some light loss, the fibre link
offers several clear advantages for instrumental stability:

\begin{itemize}
\item The optical fibres provide mechanical isolation between
  telescope and instrument,
  completely avoiding gravitational flexure and vibration.
\item Mounting the instrument away from the telescope makes it
  possible to house it in an isolated and precisely controlled
  chamber, shielding it from all environmental changes in the dome.
 \item Finally, it is well known that optical fibres tend to scramble
   the image at the fibre entrance \citep{Barden98}, reducing the
   effect of seeing and guiding variations on the illumination of the
   spectrograph.
\end{itemize}

Besides the internal absorption, which is small for fibre lengths
up to a few dozen metres, the most important source of throughput loss
in an optical fibre is caused by focal ratio degradation
(FRD). Microbendings, imperfections, and stress in the fibre scatter
the propagation angle of the rays of light, resulting in a larger
aperture of the output beam or a degraded focal ratio. FRD has a
negative effect on the resolution-throughput product of the
spectrograph, decreasing at least one of these two parameters. The FRD
properties of various fibre types have been examined in different
set-ups and several reports about the efficiency of fibres installed
on telescopes exist \citep[e.g.][]{Avila98, Ramsey88}. We think it
safe to conclude that for most fibres, FRD depends much more on the
quality of the fibre polishing, alignment, mounting, glueing,
etc. than on the fibre type, so we performed no extensive FRD testing during
the selection of the HERMES' fibres. However, fibre confectioning was done 
with outmost care to keep light loss in the fibre small.

\subsubsection{Optical configuration}
Four optical fibres are used for the HERMES spectrograph: two star
fibres mounted in the telescope interface at the focal plane of the
Mercator telescope; a calibration fibre between calibration unit and
telescope interface for illuminating the star fibres and finally a
wavelength reference fibre (WRF) that feeds wavelength calibration
light directly from the calibration unit to the spectrograph during
observations with simultaneous thorium exposures. All fibres are
multi-mode, FBP-type fibres (Polymicro, USA) without important OH$^{-}$
absorption bands in the near IR like many other fibres. The fibre
length between telescope and spectrograph is 35\,m. The sky fibres are
a low-resolution fibre (LRF) with a core diameter of 60\,$\mu$m
yielding a sky aperture of 2.15\,arcsec, and a high-resolution fibre
(HRF) with a core diameter of 80\,$\mu$m and a sky aperture of
2.5\,arcsec. It might seem contradictory that the larger diameter
fibre provides higher resolution, but only this fibre is equipped with
a slicer that effectively halves its diameter and doubles spectral
resolution. Both fibres have a 125-$\mu$m diameter cladding so that
they can be mounted easily in standard FC ferrules.

An AR-coated plano-convex micro lens (Throl Optics, Germany), glued on top of each star fibre and mounted behind a small aperture in the telescope focal plane, images the telescope pupil onto its rear surface, filling 90\% of the fibre's entrance-surface (Fig.~\ref{fig:fibre_entrance} a). At the same time, this lens reduces the $f/12$ telescope focal ratio to $f/4.9$ for HRF and $f/4.35$ for LRF. These values are reasonably well-suited to minimising focal-ratio degradation (FRD) in the fibre link, especially since at the other end of the fibres, the spectrograph accepts an aperture of $f/3.7$. Details of the fibre link optics are summarised in Table~\ref{tab:fibrelink}. It is worth noting that the telescope pupil and not the star image is formed on the fibre entrance. This allows for a much simpler optical and mechanical system.

\begin{table}
\caption{\label{tab:fibrelink}Fibre link characteristics.}
\begin{center}
\begin{tabular}{lcc} \hline\hline \rule[0mm]{0mm}{3mm}
 & LRF & HRF \\
\hline
Fibre core diameter & 60 $\mu$m & 80 $\mu$m \\
Sky aperture  & 2.15 arcsec & 2.50 arcsec \\
Aperture diameter & 150 $\mu$m & 175 $\mu$m \\
Microlens:  & & \\
\hspace{4 mm} Focal length & 0.65 mm & 0.85 mm \\
\hspace{4 mm} Lens glass type & N-LAK33 & N-BAK1 \\
\hspace{4 mm} Glass refractive index $n_{d}$ & 1.75 & 1.57 \\
\hspace{4 mm} Radius of curvature & 0.50 mm & 0.50 mm \\
\hspace{4 mm} Lens thickness & 1.15 mm & 1.35 mm \\
\hspace{4 mm} Lens diameter & 0.90 mm & 0.90 mm \\
Entrance focal ratio  & $f/4.35$ & $f/4.9$ \\
\hline
\end{tabular}
\end{center}
\end{table}

We estimate the total throughput of the HRF link, including the micro lens coupling, internal absorption, and FRD losses, to be at least 70\%. The large sky aperture results in almost negligible seeing losses in case of good (sub-arcsecond) seeing.

\subsubsection{Scrambler}
\begin{figure}
\resizebox{\hsize}{!}{\includegraphics[angle=90]{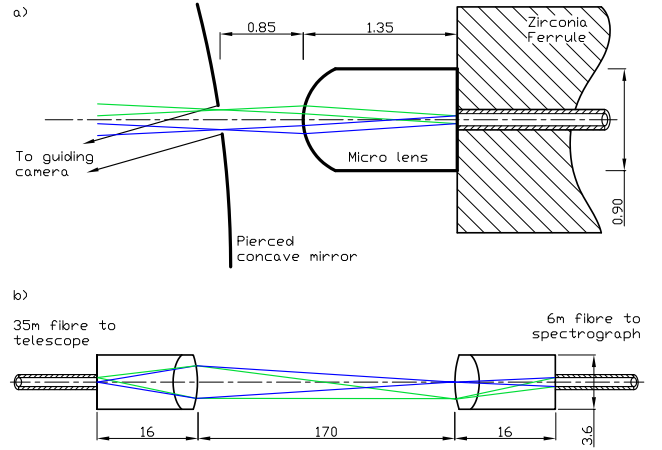}}
\caption{\label{fig:fibre_entrance} Optical fibre link: a) fibre entrance (HRF) in telescope focal plane and b) double scrambler with each fibre end imaged on to the other fibre end's pupil. Scaling has been adjusted for clarity, dimensions (mm) have been added for reference.}
\end{figure}

Although an optical fibre is an almost perfect scrambler of the spatial information at the fibre entrance, even a very long fibre does not completely remove angular information. Since we use pupil imaging, seeing variations, and guiding imprecisions mostly result in angular fluctuations. These may still induce some variation in the illumination of the echelle grating and thus introduce some error in radial-velocity measurements. To further improve illumination stability, we decided to add a double image scrambler as described by \citet{Connes85}, tested by \citet{Casse97} among others and successfully used on HARPS \citep{Pepe00} and others. A symmetric optical system in the middle of the fibre converts the far field of the first half fibre to the near field of the second half and vice versa (Fig.~\ref{fig:fibre_entrance} b). The pupil of one fibre is imaged onto the entrance of the other fibre, and thus angular non-uniformity is transformed into spatial non-uniformity that is perfectly scrambled by the second part of the fibre. In this way, the residual angular non-uniformity not scrambled by the first part of the fibre link is scrambled by the second. The scrambler consists of two identical doublets at a distance of 170\,mm (Fisba Optik, Switzerland). The exit pupil of the first fibre system fills the 60-$\mu$m core of the second fibre exactly. 

The scrambling gain (SG) is defined as the ratio between the displacement of the star in front of the input fibre end and the shift of the point spread function (PSF) on the spectrograph detector: 
\indent\[\hspace{6 mm}\mbox{SG} = \frac{d/D}{f/F}\]
where $D$ is the fibre diameter, $d$ the star shift, $f$ the shift of the PSF, and $F$ the full width at half maximum of the PSF. We measured a value of 1150 for the near field contribution of the scrambler to the SG. For comparison, the SG of the same 60-$\mu$m fibre without scrambler was measured to be only 123.

The improved illumination stability of the grating by the scrambler comes at the cost of lower throughput. The scrambler is therefore only implemented on the low-resolution fibre, optimised for high-stability radial velocity work. The sliced high-resolution fibre focuses on high efficiency and does not incorporate a scrambler. We measured the efficiency of the scrambled fibre (LRF) with respect to the HRF to lie around 70\%; however, this is not entirely due to the scrambler alone. The smaller throughput can be partly explained by the smaller LRF fibre diameter (60 versus 80\,$\mu$m).

\subsection{Telescope interface and calibration unit}
\begin{figure}
\resizebox{\hsize}{!}{\includegraphics{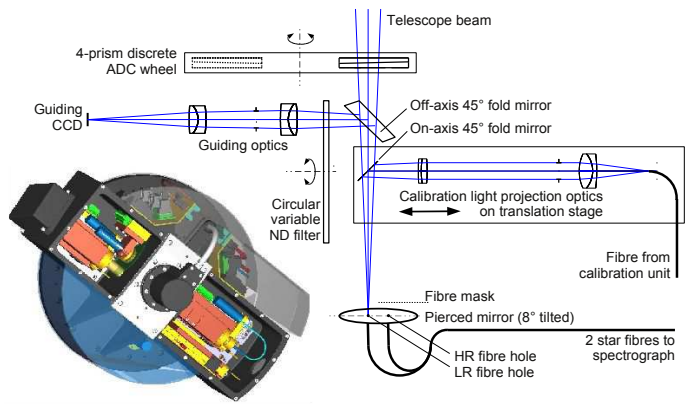}}
\caption{\label{fig:telescope_interface} Telescope interface layout and 3D view.}
\end{figure}

The telescope interface, installed at the Nasmyth A focal station of the Mercator telescope, links the telescope via the optical fibres to the spectrograph. It includes an atmospheric dispersion corrector (ADC), the fibre entrance, a fibre viewer/telescope guiding system, the calibration light projection optics, and a mask that selectively covers one or both of the star fibres (Fig.~\ref{fig:telescope_interface}). 

\subsubsection{Atmospheric dispersion corrector} 
Although the fibre entrances have a fairly large sky aperture of 2.15 or 2.5\,arcsec, HERMES would be limited to observations at zenithal angles smaller than $50^{\circ}$ if we want to avoid losing an important part of the far blue or red flux due to differential atmospheric dispersion. The HERMES atmospheric dispersion corrector provides a step-wise correction of the atmospheric dispersion by means of a set of four correctors in a selection wheel. Each corrector consists of a cemented doublet prism of N-BK7 and LLF1 glass (wedge angles: $0.55^{\circ}$, $1.85^{\circ}$, $3.60^{\circ}$ and $5.40^{\circ}$). The corrector selection matches the requirement that the length of the secondary spectrum never becomes larger than 1\,arcsec for zenithal angles up to $65^{\circ}$ and that it stays below 2\,arcsec up to $72^{\circ}$.  

\subsubsection{Acquisition and guiding}
The fibres, together with their micro lenses, are precisely mounted at the back of a polished stainless steel mirror in the focal plane of the telescope, just in front of the 175-$\mu$m and 150-$\mu$m diameter apertures (Fig.~\ref{fig:fibre_entrance} a). This concave mirror is slightly inclined ($8^{\circ}$) to reflect the image of the field via a $45^{\circ}$-fold mirror through the fibre-viewer optics to the guiding CCD. The fibre-viewer optics reduce the telescope $f/12$ focal ratio to $f/6.5$, providing a comfortable acquisition field of 4\,x\,6 arcmin on an SBIG ST-1603ME CCD camera (1020\,x\,1530 9-$\mu$m pixels). 

During acquisition, the image of the star is centred on one star-fibre hole, while a movable mask blacks out the other fibre hole. The light of the wings of the PSF that will not enter the star-fibre hole is subsequently used for telescope guiding.

\subsubsection{Calibration}
A separate calibration unit is connected to the telescope interface through a 300-$\mu$m optical fibre. The calibration light projection optics, mounted on a linear translation stage, can inject calibration light into each of the star fibres separately, thus providing optimal flat-field and wavelength calibration possibilities. The projection optics are designed to deliver a beam with exactly the same focal ratio as the telescope in order to fill the fibres in all cases with a similar light cone. 

For wavelength calibration, we use a combination of a thorium-argon lamp equipped with a red-blocking filter to cut off otherwise saturated argon lines, together with a neon lamp for additional lines in the near infrared. A double tungsten lamp is used for good  flat-field illumination. This light source consists of a normal low-power lamp providing the red part of the continuum spectrum, combined with a high-power lamp with a red-blocking filter in front of it for the blue part.

During  the simultaneous thorium mode, calibration light is fed directly to the spectrograph through the wavelength reference fibre. A variable neutral density filter in front of this fibre matches the flux from the calibration source to the integration time during this type of exposures.

\subsubsection{Observing modes}
\begin{figure}
\resizebox{\hsize}{!}{\includegraphics{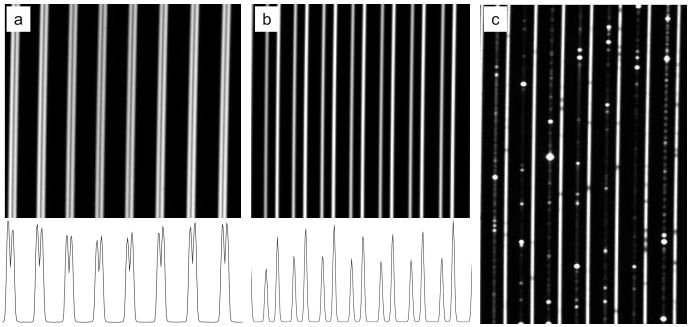}}
\caption{\label{fig:spectra_detail} Small part of raw image for a) HRF flat-field exposure, b) simultaneous LRF and reference fibre flat-field exposure, and c) LRF stellar exposure interlaced with Thorium reference exposure. The cross-cuts insets show cross-order profiles of a) the double-peaked sliced HRF and b) LRF and reference fibre (30\% smaller than LRF).}
\end{figure}

Apart of typical calibration exposures, HERMES basically offers only
two science observing modes: a) a high-resolution mode that uses
the image slicer and that is optimised for the highest possible
throughput and b) a high-precision radial velocity mode using the
simultaneous wavelength reference exposure technique. Spectral
resolution is reduced in this mode, but at the cost of lower
efficiency, spectrograph illumination is more homogenous and
stabler because of the double scrambling. As the spectrograph itself
has a completely fixed set-up, instrument configuration is done solely
through the various mechanisms in the telescope interface and
calibration unit. Fig.~\ref{fig:spectra_detail} shows a small part of
three raw images, illustrating how spectra are recorded on the
detector in different observing modes.

HERMES does not offer a sky fibre for background correction. The
limited need for sky substraction for high-resolution spectroscopy on a
1.2-m telescope does not justify the increased complexity. The
LRF mode will only be used for bright objects, and interleaving a HRF
spectrum with a sky spectrum is not possible without additional
cross dispersion because of the broad cross-order profile of the
sliced fibre. When observing a faint object with full moon, a
sky spectrum can always be taken after the science exposure.

\subsection{Detector system}
\begin{figure}
\resizebox{\hsize}{!}{\includegraphics{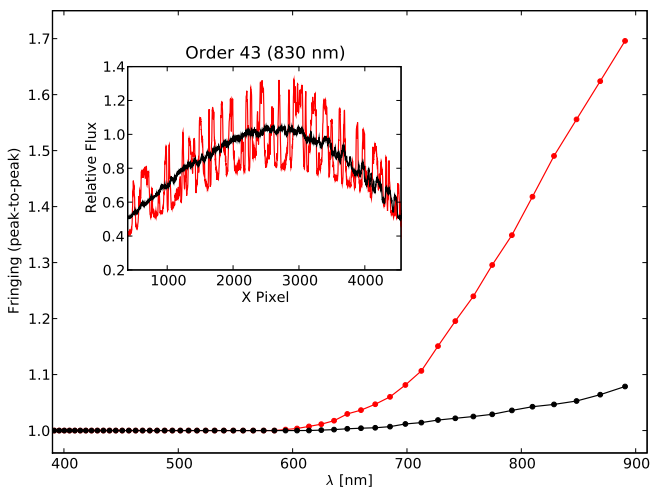}}
\caption{\label{fig:fringing} Peak-to-peak value of fringing, relative to signal flux for a standard (red) and a graded-AR coated CCD (black). The inset shows the flux in the central part of order 43  (around 830\,nm) with a standard (red) and the graded-AR coated CCD (black), illustrating the different fringing nature of both devices.}
\end{figure}

\begin{figure}
\resizebox{\hsize}{!}{\includegraphics{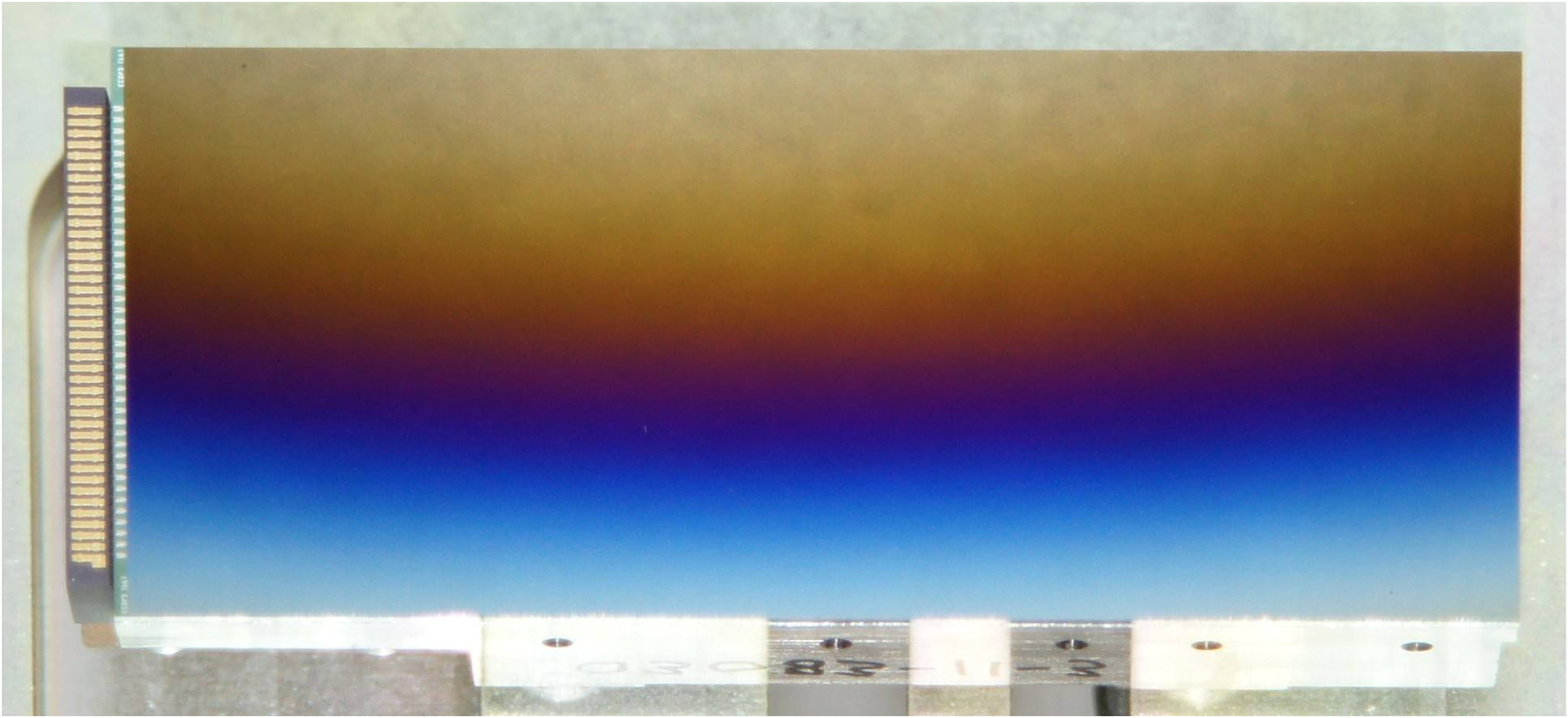}}
\caption{\label{fig:ccd} Picture of the graded-AR coated CCD. The red-sensitive part at the top of the chip looks blue because red light is absorbed while blue light is reflected.}
\vspace{3 mm}
\resizebox{\hsize}{!}{\includegraphics{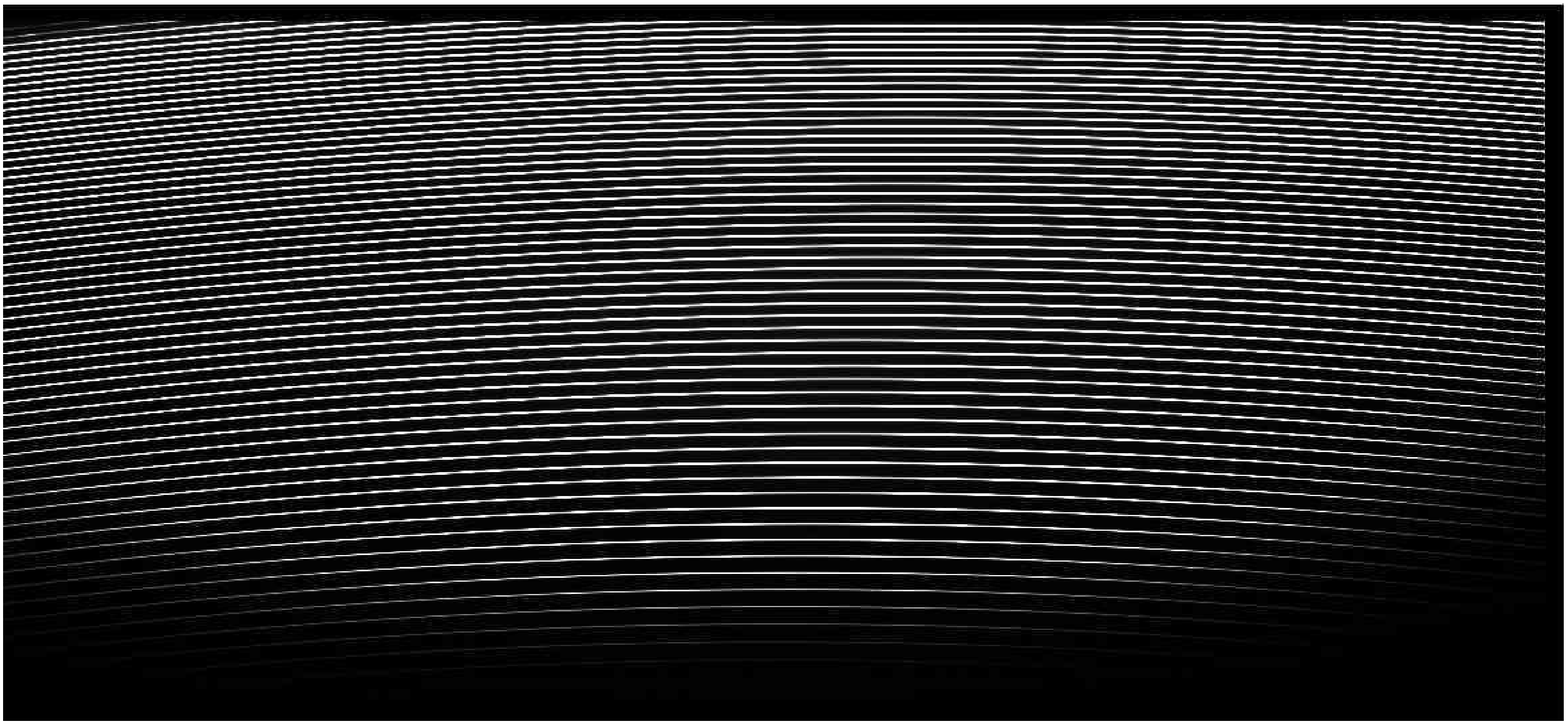}}
\caption{\label{fig:raw_image} Full-frame raw image of a flat-field spectrum in the low-resolution fibre.}
\end{figure}

HERMES uses a standard-silicon, thinned, back-illuminated e2v CCD42-90
(2048\,x\,4608 13.5-$\mu$m pixels) as detector for the
spectrograph. After a quantified analysis of spectra taken with a
deep-depletion chip, we rejected the use of such a device out of fear
of resolution degradation at the bluest wavelengths where the much
thicker silicon layer of the deep-depletion device is sensitive to
lateral charge diffusion, this despite its higher red efficiency and
strongly reduced fringing properties. To solve the fringing problems
of the thin ($\sim$15\,$\mu$m) standard silicon detector, we adopted a
novel solution developed and proposed by e2v Technologies (UK), namely
the use of a graded AR coating \citep{Kelt06}. As the HERMES
spectrograph has a completely fixed spectral format, it was possible
to optimise locally the thickness of the AR coating for the
wavelengths hitting that region of the detector. The coating
thickness follows the spectral gradient in cross-order direction, as
well as the curved pattern of the echelle orders. This of
course increases the overall detector efficiency, because for each pixel, the AR
coating is now matched to the incident monochromatic
wavelength. Moreover, fringing at NIR wavelengths is strongly reduced
because of the much lower Fresnel reflections at the CCD surface. As
can be seen from Fig.~\ref{fig:fringing}, the fringing amplitude for the reddest wavelengths of the 
device with graded AR coating is about 9 times smaller than for a similar device with astronomy broadband
AR coating. The reflection gradient of the coating can be seen in the picture of the detector in Fig.~\ref{fig:ccd}. The red-sensitive part
at the top of the CCD looks blue because red light is efficiently
absorbed and detected. As the coating is less efficient for short
wavelengths in that part of the chip, mostly blue light is
reflected. The picture of a full-frame raw image in
Fig.~\ref{fig:raw_image} shows the position of the curved echellogram
on the CCD.

The detector is mounted in a Infrared Laboratories cryostat and
cryogenically cooled by a CryoTiger closed-cycle refrigerating system
(Brooks Automation, USA). A LakeShore Model331 temperature controller
stabilises the detector temperature at 160\,K with a precision at
0.01\,K level. Detector control and read-out is done through a
standard SDSU CCD controller (Astronomical Research Cameras,
USA). Reading out all 9\,MPixels in slow mode takes 50 seconds with a
read noise of 5\,$e^{-}_{rms}$ and 30 seconds with 8\,$e^{-}_{rms}$ in
fast mode.

\subsection{Instrument control}
An industrial Linux PC, connected to the various field devices (servo
drives, remote I/O and variable power supplies) by RS-232 and RS-485
point-to-point and bus networks, controls the HERMES
instrument. Drivers have been written in Python to extend the
Mercator Observatory Control System (MOCS), currently under
development \citep{Pessemier10}. This software is
component-based, which means that HERMES could easily be integrated
within the global control system by adding components for the
detector, the instrument control, and the associated graphical user
interfaces (GUI). The instrument control software acts as a bus master
for the industrial hardware and implements the low and high level
tasks in an object-oriented way with multiple layers of
abstraction. All relevant data is published to the network, and any
interested party (such as the GUIs) can subscribe to it. Integrating
HERMES also required extending both the auto-guiding system (to allow
target acquisition and centroiding on a fibre image) and the
scheduling software.  A GUI is available for the queue scheduling of
complete nights of observations, thereby specifying the instrument
settings and exposure parameters for each target. The observer has to
supervise the system and intervene whenever needed, while the MOCS
software handles the queued observations fully automatically.

\section{Integrated data reduction pipeline}\label{sect:drs}
The data reduction pipeline performs the traditional corrections for the 
bias level, the inter-order background level, the fringing on the detector, 
and the modulation of the intensity in each spectral order (blaze function) 
and applies a pre-normalisation eliminating the global wavelength-dependency 
of the flat-field calibration system. It 
determines in a robust way the dependency of the positions of the spectral 
orders on time-dependent factors, extracts the flux in each order with 
the options of estimating the total flux in a cross-cut from the pixels free 
of cosmic rays and of weighing pixels using a cross-order flux distribution 
model. Final spectra can be represented order per order in tabular form, 
assigning a wavelength to each pixel, or resampled to bins with a size 
of $\approx 1.6$~km/s (the natural pixel size on average over each order) or 
with a fixed wavelength step, either over the whole wavelength region, over  
part of it, or just the `natural' step over each order. Depending on the 
choice, spectra may be merged over several or all orders. Care is taken 
not to include parts of orders far out of the free spectral range where 
the risk is high that systematic bias dominates random noise. A greatly 
simplified data reduction option is available for fast first-look purposes. 

\begin{figure}
\resizebox{\hsize}{8.5 cm}{\includegraphics{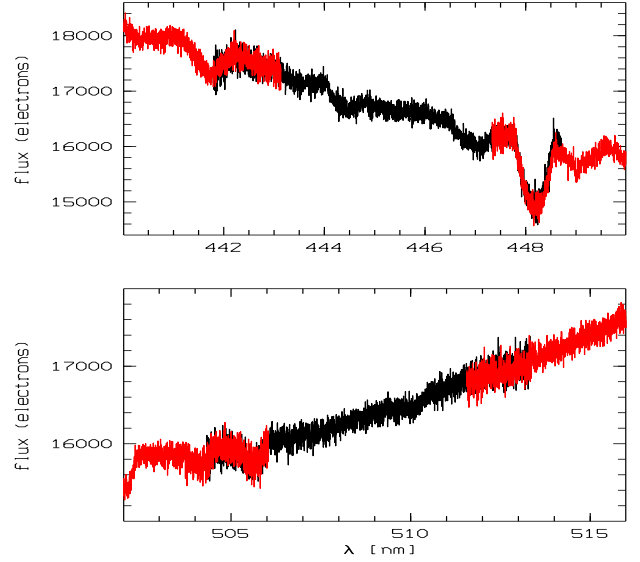}}
\caption{\label{fig:overlap}Selected regions of a late-B type binary, each panel showing a
spectral order in black and the two neighbouring orders in red.
}
\end{figure}

Presently, the data reduction pipeline works in a frame-per-frame mode, 
but the final goal is to execute a number of reduction steps in a 
differential mode to gain robustness in the temporal model of the spatial  
and wavelength geometry on the detector \citep{Hensberge07}. A detailed  
description of the data reduction procedures will be presented in a  
forthcoming paper. For the moment, we draw attention to the excellent 
agreement in the intensity level of the extracted spectra over the common 
wavelengths in subsequent orders. Fig.~\ref{fig:overlap} shows selected regions of the spectrum 
of a late-B type binary. For each order, the used wavelength range extends approximately
over the range where the intensity of the blaze is higher than
one third of its value at maximum. The extraction is
excellent and allows for detailed full spectral reconstruction, even for broad
shallow lines at the edges of spectral orders.
 
All raw and reduced frames are archived. A web-based interface for accessing 
the HERMES database is under development. The archive also provides input for 
the scheduler software when composing the observing queues.

\section{Performances}\label{sect:performance}

\subsection{Spectral resolution}
\begin{figure}
\resizebox{\hsize}{!}{\includegraphics{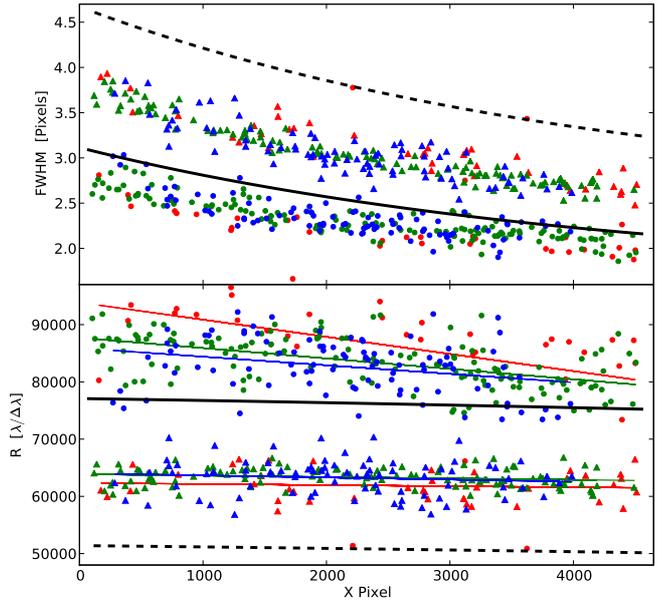}}
\caption{\label{fig:resolution} Measured line width (top panel) and spectral resolution (bottom panel) for HRF (dots) and LRF (triangles) along the detector columns. Colours correspond to wavelength for three groups of spectral orders. For comparison, the solid and dashed thick black lines show the width and resolution calculated in ZEMAX for HRF and LRF respectively.}
\end{figure}

Compared to many other spectrographs and despite a comfortably large sky aperture, HERMES really is a high-resolution spectrograph. In Fig.~\ref{fig:resolution} (bottom panel) spectral resolution is shown as derived from the measured line widths on a spectrum of thorium, argon, and neon emission lines. In LRF mode, spectral resolving power $R$\,=\,$\lambda/\Delta\lambda$ amounts to 63\,000 (4.8 km/s). This value is 23\% higher than the value of 51\,000, calculated from a ZEMAX model for a 2.15-arcsec rectangular slit. In HRF mode we measure $R$\,=\,85\,000 (3.5 km/s), 12\% higher than the calculated value of 76\,000, proving the excellent image quality of the spectrograph optics.  

\begin{figure}
\resizebox{\hsize}{6.4 cm}{\includegraphics{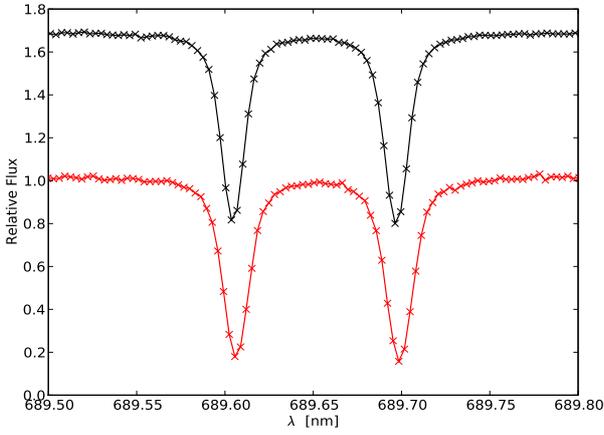}}
\caption{\label{fig:O2plot} Part of normalised spectrum with two O$_{2}$ lines measured in LRF (red) and, shown with an offset, in HRF (black).}
\end{figure}

\begin{figure*}
\includegraphics[width=18cm]{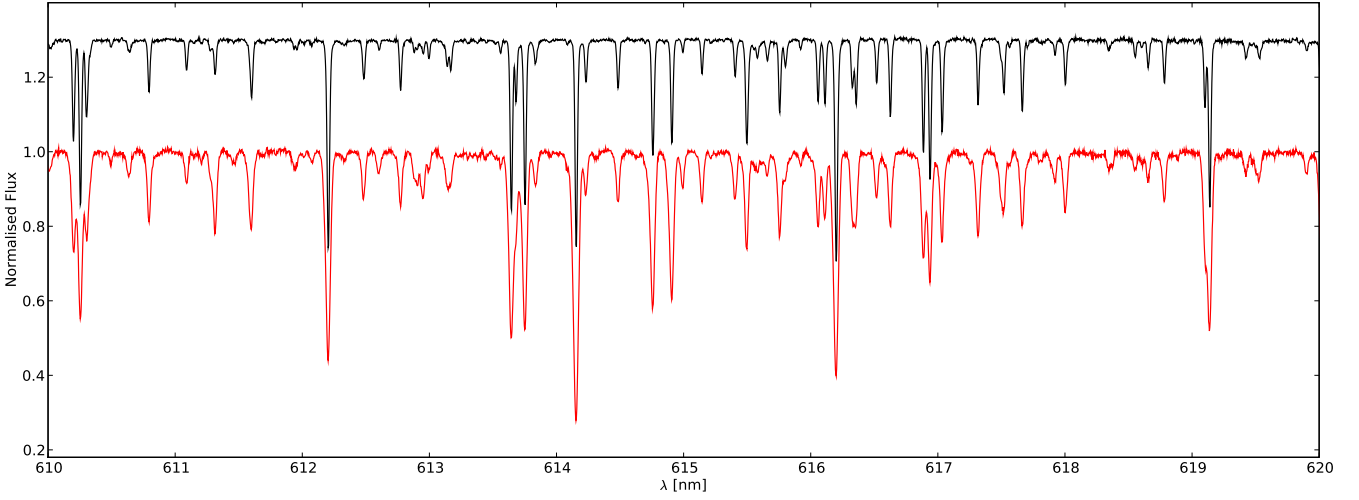}
\caption{\label{fig:spectrum} Illustrative spectra of Procyon (F5IV, black) and Polaris (F7:Ib-IIv, red). Only a small part of the spectral range is shown.}
\end{figure*}

The anamorphic magnification of the spectrograph makes the sampling
drop along the spectral orders (Fig.~\ref{fig:resolution}, top
panel). For HRF, sampling decreases from 2.7 to 2 pixels per
resolution element, and this explains the drop in HRF resolution with
X-pixel coordinate. Only for the lowest sampling values does the image
quality of the optics start becoming relevant, causing a small loss in
resolution. This effect is not noticeable for LRF because of much
higher sampling (3.8\,--\,3.1 pixels). Fig.~\ref{fig:O2plot} shows the
accurate sampling in a detailed part of the spectrum with two O$_{2}$
lines, both for HRF and LRF. Fig.~\ref{fig:spectrum} shows a larger
part of two more illustrative spectra (Procyon and Polaris).

\subsection{Efficiency}
\begin{figure}
\resizebox{\hsize}{!}{\includegraphics{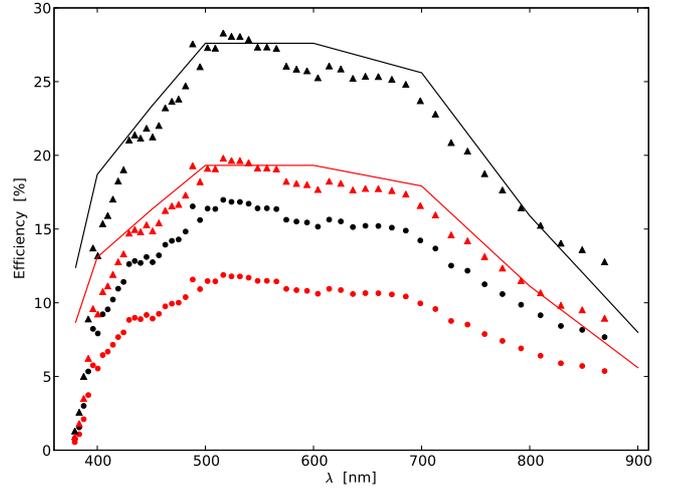}}
\caption{\label{fig:efficiency} Measured total efficiency of HERMES + Mercator telescope (dots) and efficiency of spectrograph only, including fibres and detector (triangles) for HRF (black) and LRF (red). The solid lines show the calculated spectrograph efficiency.}
\end{figure}

Fig.~\ref{fig:efficiency} shows the efficiency at blaze peak for
measurements under excellent observing conditions. Unfortunately,
according to the histogram in Fig.~\ref{fig:flux_histogram} median
observations only reach 60\% of these values. The complete system
including spectrograph and telescope has a maximum throughput of 17\%
for HRF and 12\% for LRF around 500\,nm. Taking out telescope losses
(estimated to be 40\% for three reflections), these numbers leave us
with a peak efficiency for the spectrograph alone of around 28\% and
20\%, outperforming many other first-class instruments. Apart from the
bluest wavelengths where HERMES performs below expectations, the
measured throughput corresponds quite well to calculated values
based on measurements, data sheets, and specifications and where
necessary, founded estimations of the throughput of the individual
components including fibres, mirrors, dispersers, lenses, and CCD.

With this efficiency and during good quality nights, a one hour HRF
exposure on a star of $m_{V}$\,=\,10.4 yields an S/N of 100 per pixel
around 550\,nm, decreasing to S/N\,=\,20 with $m_{V}$\,=\,13.4. For LRF
exposures, we obtain a similar S/N with stars of $m_{V}$\,=\,10.0 and 13.2.

\subsection{Stability}
The first year of operation was almost entirely devoted to science
exploitation using the HRF mode. The spectrograph and pipeline
performances are therefore best tested for this mode of
operation. Furthermore, the pipeline developments for the LRF mode have
not yet been completely finalised.

\begin{figure}
\resizebox{\hsize}{6.4 cm}{\includegraphics{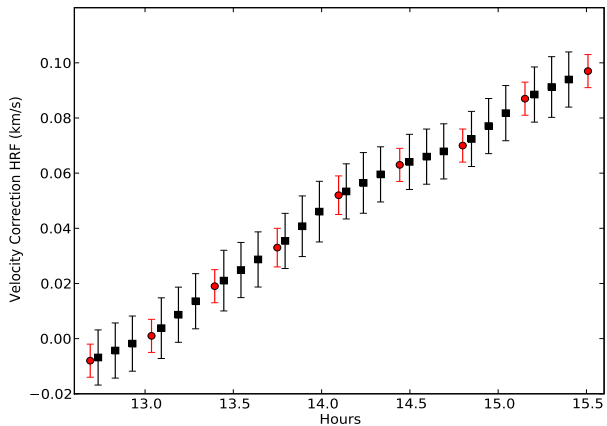}}
\caption{\label{fig:hd164922} Continous monitoring the instrumental drift of HERMES in HRF mode (red 
circles) and the interpolation during mid-exposures of the science target (black squares). }
\end{figure}

During a night, the short-term stability of the spectrograph is
strongly affected by the atmospheric pressure variations
(Sect.~\ref{sect:env_control}). To quantify the real velocity
stability of HERMES in HRF mode, we interleaved independent spectra of
the radial velocity standard HD\,164922 with wavelength calibration
frames. To correct for these short-term instrumental drifts, we
corrected the radial velocities obtained for HD\,164922 by
interpolating between the shifts found on the basis of the
neighbouring calibration frames (see Fig.~\ref{fig:hd164922}). A
standard deviation of 2.9\,m/s is found on 24 datapoints. After sigma
clipping of 2\,$\sigma$, this reduces to 2.3\,m/s.  Despite the lower
resolution of the LRF fibre, but since it is equipped with an optical
scrambler, and as the reference fibre allows for simultaneous
correction of the instrumental drift, this mode offers a slightly
higher accuracy with the standard deviation reduced to 2.5\,m/s on 45
datapoints and to 2.0\,m/s after 2\,$\sigma$ clipping.  The drawback is
the less-efficient throughput so that this mode should be reserved for
bright objects only where the errors in velocity space are dominated
by drift correction and not by photon noise, or for observations where
sampling requirements prohibit the sequential interleaving of
wavelength calibration exposures.

Although the cold head of the CryoTiger cooler does not contain any moving parts, possible vibrations caused by the gas expansion or transmitted from the compressor through the flexible gas lines (15\,m long), still raise some concern. We tried to measure these vibrations and found that, fortunately, in an experiment consisting of 300 measurements with a total time base of 4.5 hours, they are not noticeable at the detection limit of $\sigma$\,=\,70\,cm/s. The same experiment showed that also the operation of a turbo-molecur vacuum pump, connected to the detector cryostat through two metres of flexible vacuum pipe, did not cause any detectable vibrations.

The long-term stability during the first year of operation is
compromised by the continuous adjustment and optimisation of the acclimatisation
control system. In our normal mode of operation, we use only
wavelength calibration frames obtained in the evening and morning
of an observation night to calibrate a given science
frame in wavelength. Nightly we obtain spectra of IAU velocity standards. Despite the variable environmental conditions, but using
the pressure calibration of Sect.~\ref{sect:env_control}, a standard deviation of 50\,m/s is
found using all 50 exposures of the IAU standard HD\,164922 in between
2009-07-09 and 2010-06-28. 

\subsection{Scattered light and ghosts}
\begin{figure}
\resizebox{\hsize}{!}{\includegraphics{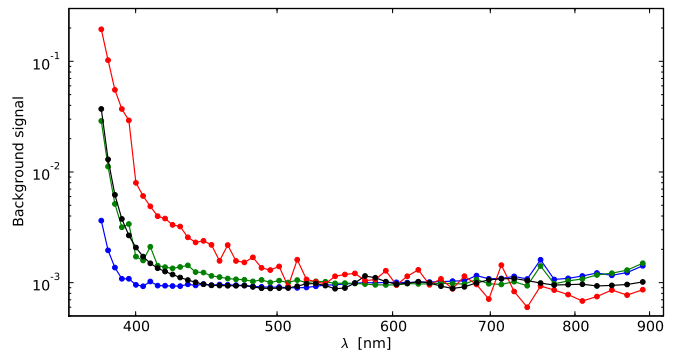}}
\caption{\label{fig:background} Ratio of signal level in the inter-order region 
to the flux in the adjacent orders for a flat field spectrum (black) and 
for an O, F, and M star (blue, green, and red curves).}
\end{figure}

Scattered light is usually an important issue in an echelle
spectrograph. Grating efficiency is typically 60-70\%, and a large part
of the lost flux ends up scattered throughout the
instrument. Nevertheless, careful design that includes efficient baffling
and use of high-quality optical components, allow HERMES to deliver
very clean spectra, comparing favourably with many other echelle
spectrographs. As can be appreciated from Fig.~\ref{fig:background},
the distribution of scattered light is very local. The inter-order
background signal fluctuates around 0.1\% of the flux in the adjacent
orders. Only when the signal level is very low (particularly at
short wavelengths) does a global straylight component become relevant.

HERMES spectra are virtually free of ghost images. In spectra of the 
emission line star P Cygni with, amongst others, a strongly saturated H$\alpha$ 
line, it turned out to be impossible to detect the presence of ghost images of 
the emission lines.

\section{Conclusion}\label{sect:conclusion}
We presented in this contribution the design, manufacturing, and
performance analyses of HERMES, a
high-resolution fibre-fed spectrograph project. 
We showed that the instrument was built according to the demanding
requirements, which resulted in an instrument of excellent stability 
and unsurpassed efficiency. The main characteristics of
our instrument are given in Table~\ref{tab:summary}. The ZEMAX
description of the optimised optical system can be found in Appendix~\ref{sect:zemax}.

\begin{table}
\caption{\label{tab:summary}HERMES main parameters.}
\begin{center}
\begin{tabular}{ll} \hline\hline \rule[0mm]{0mm}{3mm}
Spectral resolution & \\
\hspace{12 mm}HRF & 85\,000 (with slicer)\\
\hspace{12 mm}LRF & 63\,000 (interlaced with simultaneous\\
                & \hspace{9 mm} Thorium spectrum)\\
Fibre aperture & \\
\hspace{12 mm}HRF & 2.5\,arcsec  (80-$\mu$m fibre)\\
\hspace{12 mm}LRF & 2.15\,arcsec (60-$\mu$m fibre)\\
Sampling  & \\
\hspace{12 mm}HRF & 2 -- 2.7 pixels\\
\hspace{12 mm}LRF & 3.1 -- 3.8 pixels\\
Spectral range & 377 -- 900\,nm (in 55 orders)\\
Spectrograph layout & White-pupil layout on 1.2x2.4\,m bench\\
Beam diameter & 152\,mm\\
Collimator focal length & 1400\,mm (\textit{f}/9.2)\\
Echelle grating & $\theta=69.74^{\circ}$, 52.676\,gr/mm, 154x408\,mm\\
Cross disperser & 2 PBL1Y prisms, $37.4^{\circ}$ apex angle\\
Camera focal length & 475\,mm (\textit{f}/3.1,  6 lenses)\\
Detector & 2048x4608 13.5-$\mu$m pixels, thinned BI\\
         & \hspace{2 mm}CCD with graded AR coating\\
Peak efficiency & \\
\hspace{12 mm} HRF & B:\,21\%, V:\,28\%, R:\,25\%, I:\,17\% \\
\hspace{12 mm} LRF & B:\,15\%, V:\,20\%, R:\,18\%, I:\,12\% \\
Limiting magnitudes & 1 hour exposure\\
\hspace{12 mm} S/N$_{V}$\,=\,20 & $m_{V}=13.4$ (HRF), $m_{V}=13.2$ (LRF)\\
\hspace{12 mm} S/N$_{V}$\,=\,100 &$m_{V}=10.4$ (HRF), $m_{V}=10.0$ (LRF)\\
Radial velocity error & $\sim$ 2.5 m/s (short term)\\
\hline
\end{tabular}
\end{center}
\end{table}

After commissioning on the Mercator telescope in April 2009, HERMES
immediately became the work-horse instrument. During the first year of
operations, this instrument obtained about 7300 spectra on the sky,
which makes clear that HERMES also fulfills its robust operational
requirements. HERMES is open to the community at large through
collaboration with the consortium members and through the Spanish CAT
time allocation programme, available for all telescopes at the Roque de los
Muchachos observatory.

\begin{acknowledgements}
The HERMES project and team acknowledge support from the Fund for Scientific Research of
Flanders (FWO) under the grant G.0472.04, from
the Research Council of K.U.Leuven under grant GST-B4443,
from the Fonds National de la Recherche Scientifique under contracts IISN\,4.4506.05 and FRFC\,2.4533.09,
and financial support from Lotto (2004) assigned to the Royal Observatory of 
Belgium to contribute to the hardware of the spectrograph. We
would like to express our appreciation and gratitude to Francesco Pepe
(Geneva Observatory, Switzerland),  Otmar
Stahl (Heidelberg, Germany), and Ramon Garc\'{i}a Lopez
(IAC, Spain), who served as experts in the review committee, and to Jos\'{e} Luis Rasilla (IAC, Spain) for his helpful comments on the optical design.
\end{acknowledgements}

\bibliographystyle{aa}

\begin{thebibliography}{23}
\expandafter\ifx\csname natexlab\endcsname\relax\def\natexlab#1{#1}\fi

\bibitem[{{Aerts} {et~al.}(2010){Aerts}, {Christensen-Dalsgaard}, \&
  {Kurtz}}]{aerts10}
{Aerts}, C., {Christensen-Dalsgaard}, J., \& {Kurtz}, D.~W. 2010,
  {Asteroseismology}, ed. {Aerts, C., Christensen-Dalsgaard, J., \& Kurtz,
  D.~W.}

\bibitem[{{Avila}(1998)}]{Avila98}
{Avila}, G. 1998, in Astronomical Society of the Pacific Conference Series,
  Vol. 152, Fiber Optics in Astronomy III, ed. {S.~Arribas, E.~Mediavilla, \&
  F.~Watson}, 44--+

\bibitem[{{Baranne}(1988)}]{Baranne88}
{Baranne}, A. 1988, in ESO Conference on Very Large Telescopes and their
  Instrumentation, Vol. 2, p. 1195 - 1206, Vol.~2, 1195--1206

\bibitem[{{Baranne} {et~al.}(1996){Baranne}, {Queloz}, {Mayor}, {Adrianzyk},
  {Knispel}, {Kohler}, {Lacroix}, {Meunier}, {Rimbaud}, \& {Vin}}]{Baranne96}
{Baranne}, A., {Queloz}, D., {Mayor}, M., {et~al.} 1996, \aaps, 119, 373

\bibitem[{{Barden}(1998)}]{Barden98}
{Barden}, S.~C. 1998, in Astronomical Society of the Pacific Conference Series,
  Vol. 152, Fiber Optics in Astronomy III, ed. {S.~Arribas, E.~Mediavilla, \&
  F.~Watson}, 14--+

\bibitem[{{Blomme}(2009)}]{blomme09}
{Blomme}, R. 2009, ArXiv e-prints

\bibitem[{{Casse} \& {Vieira}(1997)}]{Casse97}
{Casse}, M. \& {Vieira}, F. 1997, in Society of Photo-Optical Instrumentation
  Engineers (SPIE) Conference Series, Vol. 2871, Society of Photo-Optical
  Instrumentation Engineers (SPIE) Conference Series, ed. {A.~L.~Ardeberg},
  1187--1196

\bibitem[{{Connes}(1985)}]{Connes85}
{Connes}, P. 1985, \apss, 110, 211

\bibitem[{{Dekker} {et~al.}(2000){Dekker}, {D'Odorico}, {Kaufer}, {Delabre}, \&
  {Kotzlowski}}]{Dekker00}
{Dekker}, H., {D'Odorico}, S., {Kaufer}, A., {Delabre}, B., \& {Kotzlowski}, H.
  2000, in Society of Photo-Optical Instrumentation Engineers (SPIE)
  Conference, Vol. 4008, Society of Photo-Optical Instrumentation Engineers
  (SPIE) Conference Series, ed. {M.~Iye \& A.~F.~Moorwood}, 534--545

\bibitem[{{Hensberge}(2007)}]{Hensberge07}
{Hensberge}, H. 2007, in Astronomical Society of the Pacific Conference Series,
  Vol. 364, The Future of Photometric, Spectrophotometric and Polarimetric
  Standardization, ed. {C.~Sterken}, 275--+

\bibitem[{{Hensberge} {et~al.}(2008){Hensberge}, {Iliji{\'c}}, \&
  {Torres}}]{hensberge08}
{Hensberge}, H., {Iliji{\'c}}, S., \& {Torres}, K.~B.~V. 2008, \aap, 482, 1031

\bibitem[{{Jorissen} {et~al.}(2009){Jorissen}, {Frankowski}, {Famaey}, \& {van
  Eck}}]{jorissen09}
{Jorissen}, A., {Frankowski}, A., {Famaey}, B., \& {van Eck}, S. 2009, \aap,
  498, 489

\bibitem[{{Kaufer}(1998)}]{Kaufer98b}
{Kaufer}, A. 1998, in Astronomical Society of the Pacific Conference Series,
  Vol. 152, Fiber Optics in Astronomy III, ed. {S.~Arribas, E.~Mediavilla, \&
  F.~Watson}, 337--+

\bibitem[{{Kaufer} \& {Pasquini}(1998)}]{Kaufer98}
{Kaufer}, A. \& {Pasquini}, L. 1998, in Society of Photo-Optical
  Instrumentation Engineers (SPIE) Conference Series, Vol. 3355, Society of
  Photo-Optical Instrumentation Engineers (SPIE) Conference Series, ed.
  {S.~D'Odorico}, 844--854

\bibitem[{{Kelt} {et~al.}(2006){Kelt}, {Harris}, {Jorden}, \&
  {Tulloch}}]{Kelt06}
{Kelt}, A., {Harris}, A., {Jorden}, P., \& {Tulloch}, S. 2006, in Scientific
  Detectors for Astronomy 2005, ed. {J.~E.~Beletic, J.~W.~Beletic, \&
  P.~Amico}, 369--+

\bibitem[{{Lehmann} {et~al.}(2010){Lehmann}, {Vitrichenko}, {Bychkov},
  {Bychkova}, \& {Klochkova}}]{lehmann10}
{Lehmann}, H., {Vitrichenko}, E., {Bychkov}, V., {Bychkova}, L., \&
  {Klochkova}, V. 2010, \aap, 514, A34+

\bibitem[{{Lindzen}(1979)}]{Lindzen79}
{Lindzen}, R.~S. 1979, Annual Review of Earth and Planetary Sciences, 7, 199

\bibitem[{{Masseron} {et~al.}(2010){Masseron}, {Johnson}, {Plez}, {van Eck},
  {Primas}, {Goriely}, \& {Jorissen}}]{masseron10}
{Masseron}, T., {Johnson}, J.~A., {Plez}, B., {et~al.} 2010, \aap, 509, A93+

\bibitem[{{Pepe} {et~al.}(2000){Pepe}, {Mayor}, {Delabre}, {Kohler}, {Lacroix},
  {Queloz}, {Udry}, {Benz}, {Bertaux}, \& {Sivan}}]{Pepe00}
{Pepe}, F., {Mayor}, M., {Delabre}, B., {et~al.} 2000, in Society of
  Photo-Optical Instrumentation Engineers (SPIE) Conference, Vol. 4008, Society
  of Photo-Optical Instrumentation Engineers (SPIE) Conference Series, ed.
  {M.~Iye \& A.~F.~Moorwood}, 582--592

\bibitem[{{Pessemier} {et~al.}(2010){Pessemier}, {Raskin}, {Prins}, {Saey},
  {Merges}, {Padilla}, {van Winckel}, \& {Waelkens}}]{Pessemier10}
{Pessemier}, W., {Raskin}, G., {Prins}, S., {et~al.} 2010, in Society of
  Photo-Optical Instrumentation Engineers (SPIE) Conference Series, Vol. 7740,
  Society of Photo-Optical Instrumentation Engineers (SPIE) Conference Series,
  77403B--77403B--10

\bibitem[{{Ramsey}(1988)}]{Ramsey88}
{Ramsey}, L.~W. 1988, in Astronomical Society of the Pacific Conference Series,
  Vol.~3, Fiber Optics in Astronomy, ed. {S.~C.~Barden}, 26--39

\bibitem[{{Reyniers} {et~al.}(2007){Reyniers}, {Abia}, {Van Winckel}, {Lloyd
  Evans}, {Decin}, {Eriksson}, \& {Pollard}}]{reyniers07a}
{Reyniers}, M., {Abia}, C., {Van Winckel}, H., {et~al.} 2007, \aap, 461, 641

\bibitem[{{Van Winckel} {et~al.}(2009){Van Winckel}, {Lloyd Evans}, {Briquet},
  {De Cat}, {Degroote}, {De Meester}, {De Ridder}, {Deroo}, {Desmet},
  {Drummond}, {Eyer}, {Groenewegen}, {Kolenberg}, {Kilkenny}, {Ladjal},
  {Lefever}, {Maas}, {Marang}, {Martinez}, {{\O}stensen}, {Raskin}, {Reyniers},
  {Royer}, {Saesen}, {Uytterhoeven}, {Vanautgaerden}, {Vandenbussche}, {van
  Wyk}, {Vu{\v c}kovi{\'c}}, {Waelkens}, \& {Zima}}]{vanwinckel09}
{Van Winckel}, H., {Lloyd Evans}, T., {Briquet}, M., {et~al.} 2009, \aap, 505,
  1221

\end{thebibliography}

\appendix
\section{ZEMAX optical description}\label{sect:zemax}
Table~\ref{tab:surface_data} summarises the \textit{as built} ZEMAX data of all optical surfaces of the HERMES spectrograph. The optical system has been optimised by considering the measured values of the refractive indices of the optical glasses. This means that the use of standard refractive index values results in a slightly suboptimal image quality for the system that is described here.

\begin{table*}
\caption{\label{tab:surface_data}ZEMAX optical surface data.}
\begin{center}
\begin{tabular}{rccccccl} \hline\hline \rule[0mm]{0mm}{3mm}
Surface & Type &        Radius &     Thickness &               Glass &     Diameter &         Conic &  Comment\\
\hline
 OBJECT & STANDARD &  Infinity &         0 &              SILICA &         0.62 &             0 &  FIBRE EXIT\\
   1 & STANDARD &  Infinity &        12 &              SILICA &         0.62 &             0 &  FN DOUBLET \\
   2 & STANDARD &     3.157 &       3.5 &             S-FPL53 &          5.6 &             0 &   \\  
   3 & STANDARD &    $-$6.048 &        24 &                     &          5.6 &             0 &   \\  
 STOP & STANDARD &  Infinity &      39.5 &                     &          4.6 &             0 &   \\  
   5 & STANDARD &     28.04 &      3.05 &             S-BAL12 &           12 &             0 &  FN TRIPLET \\
   6 & STANDARD &    14.009 &      5.83 &             S-FPL53 &           12 &             0 &   \\
   7 & STANDARD &   $-$14.009 &      3.06 &             S-BAL12 &           12 &             0 &   \\
   8 & STANDARD &    $-$28.04 &     59.51 &                     &          9.1 &             0 &   \\
   9 & STANDARD &  Infinity &      12.5 &              SILICA &           25 &             0 &  SLICER \\
  10 & STANDARD &  Infinity &      12.5 &              SILICA &           25 &             0 &  SLICER\\
  11 & STANDARD &  Infinity &   1401.48 &                     &        2.424 &             0 &   \\  
  12 & COORDBRK \tablefootmark{a} &         - &         0 &                     &            - &             - &  OFF-AXIS ANGLE \\
  13 & STANDARD  \tablefootmark{b}&  $-$2802.96 &  $-$1401.48 &              MIRROR &          660 &            $-$1 &  MAIN COLLIMATOR \\
  14 & COORDBRK \tablefootmark{c} &         - &         0 &                     &            - &             - &  ECHELLE TILT \\
  15 & COORDBRK \tablefootmark{d} &         - &         0 &                     &            - &             - &  LITTROW ANGLE \\
  16 & DGRATING  \tablefootmark{e} &  Infinity &         0 &              MIRROR &       444.71 &             0 &  ECHELLE GRATING\\
  17 & COORDBRK  \tablefootmark{f}&         - &         0 &                     &            - &             - &   \\
  18 & COORDBRK \tablefootmark{g} &         - &   1401.48 &                     &            - &             - &   \\  
  19 & STANDARD \tablefootmark{h} &  $-$2802.96 &     $-$1300 &              MIRROR &          660 &            $-$1 &  MAIN COLLIMATOR \\
  20 & STANDARD &  Infinity &   1504.11 &              MIRROR &       114.80 &             0 &  FLAT FOLD MIRROR \\
  21 & STANDARD  \tablefootmark{i} &  $-$2802.96 &     $-$1280 &              MIRROR &          660 &            $-$1 &  TRANSFER COLLIMATOR \\
  22 & COORDBRK  \tablefootmark{j}&         - &         0 &                     &            - &             - &  PRISM ANGLE OF INCIDENCE \\
  23 & STANDARD &  Infinity &         0 &               PBL1Y &          196 &             0 &  CD PRISM1 \\
  24 & COORDBRK \tablefootmark{k} &         - &       $-$71 &                     &            - &             - &  PRISM HALF APEX ANGLE \\
  25 & COORDBRK  \tablefootmark{l}&         - &         0 &                     &            - &             - &  PRISM HALF APEX ANGLE \\
  26 & STANDARD &  Infinity &         0 &                     &          196 &             0 &  \\
  27 & COORDBRK  \tablefootmark{l}&         - &      $-$106 &                     &            - &             - &  PRISM ANGLE OF INCIDENCE\\
  28 & COORDBRK  \tablefootmark{l}&         - &         0 &                     &            - &             - &  PRISM ANGLE OF INCIDENCE\\
  29 & STANDARD &  Infinity &         0 &               PBL1Y &          196 &             0 &  CD PRISM2\\
  30 & COORDBRK \tablefootmark{k} &         - &       $-$71 &                     &            - &             - &  PRISM HALF APEX ANGLE\\
  31 & COORDBRK  \tablefootmark{k}&         - &         0 &                     &            - &             - &  PRISM HALF APEX ANGLE\\
  32 & STANDARD &  Infinity &         0 &                     &          196 &             0 &  \\
  33 & COORDBRK  \tablefootmark{l}&         - &       $-$75 &                     &            - &             - &  PRISM ANGLE OF INCIDENCE\\
  34 & STANDARD &   $-$316.45 &    $-$21.19 &             S-FPL53 &          172 &             0 &  CAMERA LENS1\\
  35 & STANDARD &   1216.54 &    $-$1.053 &                     &          172 &             0 &  \\
  36 & STANDARD &   $-$331.92 &   $-$16.127 &             S-BAL42 &          172 &             0 &  LENS2\\
  37 & STANDARD &   $-$165.26 &   $-$35.949 &             S-FPL53 &          164 &             0 &  LENS3\\
  38 & STANDARD &    437.42 &   $-$15.105 &             S-LAL12 &          164 &             0 &  LENS4\\
  39 & STANDARD &   $-$580.77 &  $-$311.064 &                     &          164 &             0 &  \\
  40 & STANDARD &   $-$326.08 &   $-$21.277 &              PBL25Y &          154 &             0 &  LENS5\\
  41 & STANDARD &   1700.88 &   $-$199.41 &                     &          154 &             0 &  ADJUST FOCUS\\
  42 & STANDARD &    133.26 &     $-$14.1 &              SILICA &           88 &             0 &  FIELD LENS \\
  43 & TOROIDAL &       224 &      $-$5.3 &              VACUUM &           80 &             0 &  CYLINDRICAL SURFACE\\
 IMAGE & STANDARD &  Infinity &           &                     &         41 &             0 &  CCD\\
\hline
\end{tabular}
\end{center}

\tablefoot{
\tablefoottext{a}{Tilt about Y: 6.32} 
\tablefoottext{b}{Decenter X before surface: 155, Decenter X after surface: $-$155 } 
\tablefoottext{c}{Tilt about Y: 0.8} 
\tablefoottext{d}{Tilt about X: 69.74} 
\tablefoottext{e}{Lines\,/\,$\mu$m: 0.052676, Diffraction order: 40--94} 
\tablefoottext{f}{Tilt about X: $-$69.74} 
\tablefoottext{g}{Tilt about Y: $-$0.8} 
\tablefoottext{h}{Decenter X before surface: 155} 
\tablefoottext{i}{Decenter X after surface: 158} 
\tablefoottext{j}{Tilt about Y: $-$29.545} 
\tablefoottext{k}{Tilt about Y: 18.7} 
\tablefoottext{l}{Tilt about Y: $-$29.545} 
}
\end{table*}

\end{document}